\documentclass[12pt]{revtex4}
\usepackage{amssymb,epsf}
\usepackage{latexsym}

\begin{document}

\title{Topological Black Holes in Brans-Dicke-Maxwell Theory }
\author{A. Sheykhi$^{1,2}$\footnote{sheykhi@mail.uk.ac.ir} and  H. Alavirad$^{1}$}
\address{$^1$Department of Physics, Shahid Bahonar University, P.O. Box 76175, Kerman, Iran\\
         $^2$Research Institute for Astronomy and Astrophysics of Maragha (RIAAM), Maragha, Iran}
\begin{abstract}
We derive a new analytic solution of $(n+1)$-dimensional $(n\geq
4)$ Brans-Dikce-Maxwell theory in the presence of a potential for
the scalar field, by applying a conformal transformation to the
dilaton gravity theory. These solutions describe topological
charged black holes with unusual asymptotics. We obtain the
conserved and thermodynamic quantities through the use of the
Euclidean action method. We also study the thermodynamics of the
solutions and verify that the conserved and thermodynamic
quantities of the solutions satisfy the first law of black hole
thermodynamics.


\end{abstract}

 \maketitle
\section{Introduction\label{I}}

Among the alternative theories of general relativity, perhaps, the
most well-known theory is the scalar-tensor theory which was
pioneered several decades ago by Brans and Dicke \cite{BD}, who
sought to incorporate Mach's principle into gravity. Compared to
Einstein's general relativity, Brans-Dicke (BD) theory describes
the gravitation in terms of the metric as well as scalar fields
and accommodates both Mach's principle and Dirac's large number
hypothesis as new ingredients. Although BD theory has passed all
the possible observational tests \cite{Will}, however, the
singularity problem remains yet in this theory. In recent years
this theory got a new impetus as it arises naturally as the low
energy limit of many theories of quantum gravity such as the
supersymmetric string theory or the Kaluza-Klein theory. Besides,
recent observations show that at the present epoch, our Universe
expands with acceleration instead of deceleration along the scheme
of standard Friedmann models and since general relativity could
not describe such Universe correctly, cosmologists have attended
to alternative theories of gravity such as BD theory. Due to
highly nonlinear character of BD theory, a desirable pre-requisite
for studying strong field situation is to have knowledge of exact
explicit solutions of the field equations. And as black holes are
very important both in classical and quantum gravity, many authors
have investigated various aspects of them in BD theory
\cite{Sen,Sen2,Sen3}. It turned out that the dynamic scalar field
in the BD theory plays an important role in the process of
collapse and critical phenomenon. The first four-dimensional black
hole solutions of BD theory was obtained by Brans in four classes
\cite{Brans}. It has been shown that among these four classes of
the static spherically symmetric solutions of the vacuum BD theory
of gravity only two are really independent, and only one of them
is permitted for all values of $\omega$. It has been proved that
in four dimensions, the stationary and vacuum BD solution is just
the Kerr solution with constant scalar field everywhere
\cite{Hawking}. It has been shown that the charged black hole
solution in four-dimensional Brans-Dicke-Maxwell (BDM) theory is
just the Reissner-Nordstrom solution with a constant scalar field,
however, in higher dimensions, one obtains the black hole
solutions with a nontrivial scalar field \cite{Cai1}. This is
because the stress energy tensor of Maxwell field is not traceless
in the higher dimensions and the action of Maxwell field is not
invariant under conformal transformations. Accordingly, the
Maxwell field can be regarded as the source of the scalar field in
the BD theory \cite{Cai1}. Other studies on black hole solutions
in BD theory have been carried out in \cite{Kim,Deh1,Gao,Kim2}.

On another front, it has been realized that in four dimensions the
topology of the event horizon of an asymptotically flat stationary
black hole is uniquely determined to be the two-sphere $S^2$
\cite{Haw1,Haw2}. The ``topological censorship theorem" of
Friedmann, Schleich and Witt indicates the impossibility of non
spherical horizons \cite{FSW1,FSW2}. However, when the asymptotic
flatness of spacetime is violated, there is no fundamental reason
to forbid the existence of static or stationary black holes with
nontrivial topologies. It was shown that for asymptotically AdS
spacetime, in the four-dimensional Einstein-Maxwell theory, there
exist black hole solutions whose event horizons may have zero or
negative constant curvature and their topologies are no longer the
two-sphere $S^2$. The properties of these black holes are quite
different from those of black holes with usual spherical topology
horizon, due to the different topological structures of the event
horizons. Besides, the black hole thermodynamics is drastically
affected by the topology of the event horizon. It was argued that
the Hawking-Page phase transition \cite{Haw3} for the
Schwarzschild-AdS black hole does not occur for locally AdS black
holes whose horizons have vanishing or negative constant
curvature, and they are thermally stable \cite{Birm}. The studies
on the topological black holes have been carried out extensively
in many aspects \cite{Lem,Cai2,Vanzo,Bril1,Cai3,Cai4,
Shey1,shey2,Cri,Cri2,Cri3,MHD,jplemos,huang,rgcai0,rbmann,Ban,rgcai1,rgcai2}.
In this paper, we would like to study the topological black hole
solutions in $(n+1)$-dimensional BDM theory for an arbitrary value
of $\omega$ and investigate their properties.

In the next section, we review the basic equations and the
conformal transformation between the action of the dilaton gravity
theory and the BD theory. In section \ref{III}, we construct
charged topological black hole solutions in BDM theory and
investigate their properties. In section \ref{IV}, we study the
thermodynamical properties of the solutions and calculate the
conserved quantities through the use of the Euclidean action
method. The last section is devoted to summary and discussion.

\section{ Basic equations and Conformal Transformations}\label{II}
The action of the $(n+1)$-dimensional Brans-Dicke-Maxwell theory
with one scalar field $\Phi$ and a self-interacting potential
$V(\Phi)$ can be written as
\begin{equation}
I_{G}=-\frac{1}{16\pi}\int_{\mathcal{M}}
d^{n+1}x\sqrt{-g}\left(\Phi {R}-\frac{\omega}{\Phi}(\nabla\Phi)^2
-V(\Phi)-F_{\mu \nu}F^{\mu \nu}\right)-
\frac{1}{8\pi}\int{d^{n}x\sqrt{h}\Phi K},\label{act1}
\end{equation}
where ${R}$ is the scalar curvature, $V(\Phi )$ is a potential for
the scalar field $\Phi $, $F_{\mu \nu }=\partial _{\mu }A_{\nu
}-\partial _{\nu }A_{\mu }$ is the electromagnetic field tensor,
and $A_{\mu }$ is the electromagnetic potential. The factor
$\omega$ is the coupling constant. The last term in Eq.
(\ref{act1}) is the Gibbons-Hawking surface term. It is required
for the variational principle to be well-defined. The factor $K$
represents the trace of the extrinsic curvature for the boundary
and $h$ is the induced metric on the boundary. The equations of
motion can be obtained by varying the action (\ref{act1}) with
respect to the gravitational field $g_{\mu \nu }$, the scalar
field $\Phi $ and the gauge field $A_{\mu }$ which yields the
following field equations
\begin{eqnarray}
&&G_{\mu
\nu}=\frac{\omega}{\Phi^2}\left(\nabla_{\mu}\Phi\nabla_{\nu}\Phi-\frac{1}{2}g_{\mu
\nu}(\nabla\Phi)^2\right)
-\frac{V(\Phi)}{2\Phi}g_{\mu \nu}+\frac{1}{\Phi}\left(\nabla_{\mu}\nabla_{\nu}\Phi-g_{\mu \nu}\nabla^2\Phi\right)\nonumber \\
&&+\frac{2}{\Phi}\left(F_{\mu \lambda}F_{ \nu}^{\
\lambda}-\frac{1}{4}F_{\rho \sigma}F^{\rho
\sigma}g_{\mu \nu}\right), \label{Eq1}\\
&&\nabla^2\Phi=-\frac{n-3}{2(n-1)\omega+2n}F^2+\frac{1}{2(n-1)\omega+2n}\left((n-1)\Phi\frac{dV(\Phi)}{d\Phi}
-(n+1)V(\Phi)\right),\label{Eq2} \\
&&\nabla_{\mu}F^{\mu \nu}=0, \label{Eq3}
\end{eqnarray}
where $G_{\mu \nu}$ and $\nabla$ are, respectively, the Einstein
tensor and covariant differentiation in the spacetime metric
$g_{\mu \nu}$. It is apparent that the right hand side of Eq.
(\ref{Eq1}) includes the second derivatives of the scalar field,
so it is hard to solve the field equations (\ref{Eq1})-(\ref{Eq3})
directly. We can remove this difficulty by a conformal
transformation. Indeed, the BDM theory (\ref{act1}) can be
transformed into the Einstein-Maxwell theory with a minimally
coupled scalar field via the conformal transformation
\begin{eqnarray}
&&\bar{g}_{\mu \nu}=\Phi^{\frac{2}{n-1}}g_{\mu \nu}, \nonumber \\
&&\bar{\Phi}=\frac{n-3}{4\alpha}\ln \Phi, \label{conf}
\end{eqnarray}
where
\begin{equation}
\alpha=\frac{n-3}{\sqrt{4(n-1)\omega+4n}}. \label{6}
\end{equation}
Using this conformal transformation, the action (\ref{act1})
transforms to
\begin{equation}
\bar{I}_{G}=-\frac{1}{16\pi}\int_{\mathcal{M}}
d^{n+1}x\sqrt{-\bar{g}}\left({\bar{R}}-\frac{4}{n-1}(\bar{\nabla}\
\bar{\Phi})^2-\bar{V}(\bar{\Phi})-e^{-\frac{4\alpha\bar{\Phi}}{n-1}}\bar{F}_{\mu
\nu}\bar{F}^{\mu \nu}\right), \label{act2}
\end{equation}
where ${\bar{R}}$ and $\bar{\nabla}$ are, respectively, the Ricci
scalar and covariant differentiation in the spacetime metric
$\bar{g}_{\mu \nu}$, and $\bar{V}(\bar{\Phi})$ is
\begin{equation}
\bar{V}(\bar{\Phi})=\Phi^{-\frac{n+1}{n-1}}V(\Phi).\label{8}
\end{equation}
This action is just the action of the $(n+1)$-dimensional
Einstein-Maxwell-dilaton gravity, where $\bar{\Phi}$ is the
dilaton field and $\bar{V}(\bar{\Phi})$ is a potential for
$\bar{\Phi}$. $\alpha $ is an arbitrary constant governing the
strength of the coupling between the dilaton and the Maxwell
field. Varying the action (\ref{act2}), we can obtain equations of
motion
\begin{eqnarray}
&&\bar{{R}}_{\mu
\nu}=\frac{4}{n-1}\left(\bar{\nabla}_{\mu}\bar{\Phi}\bar{\nabla}
_{\nu}\bar{\Phi}+\frac{1}{4}\bar{V}(\bar{\Phi})\bar{g}_{\mu
\nu}\right)+ 2e^{\frac{-4\alpha\bar{\Phi}}{n-1}}\left(\bar{F}_{\mu
\lambda}\bar{F}_{\nu}^{ \ \lambda} -\frac{1}{2(n-1)}\bar{F}_{\rho
\sigma}\bar{F}^{\rho \sigma}\bar{g}_{\mu \nu}\right), \label{Eqd1}\\
&&
\bar{\nabla}^2\bar{\Phi}=\frac{n-1}{8}\frac{\partial\bar{V}}{\partial\bar{\Phi}}
-\frac{\alpha}{2}e^{\frac{-4\alpha\bar{\Phi}}{n-1}}\bar{F}_{\rho
\sigma}\bar{F}^{\rho \sigma},\label{Eqd2}\\
&&
\bar{\nabla}_{\mu}\left(e^{\frac{-4\alpha\bar{\Phi}}{n-1}}\bar{F}^{\mu
\nu}\right)=0. \label{Eqd3}
\end{eqnarray}
Comparing Eqs. (\ref{Eq1})-(\ref{Eq3}) with Eqs.
(\ref{Eqd1})-(\ref{Eqd3}), we find that if $\left(\bar{g}_{\mu
\nu},\bar{F}_{\mu \nu},\bar{\Phi}\right)$ is the solution of Eqs.
(\ref{Eq1})-(\ref{Eq3}) with potential $\bar{V}(\bar{\Phi})$, then
\begin{equation}\label{conform2}
\left[{g}_{\mu \nu},{F}_{\mu
\nu},{\Phi}\right]=\left[\exp\left({\frac{-8\alpha
\bar{\Phi}}{(n-1)(n-3)}}\right)\bar{g}_{\mu \nu},\bar{F}_{\mu
\nu},\exp\left({\frac{4\alpha \bar{\Phi}}{n-3}}\right)\right],
\end{equation}
is the solution of Eqs. (\ref{Eqd1})-(\ref{Eqd3}) with potential
$V(\Phi)$.
\section{Topological black holes in BDM theory\label{III}}
The solutions of the field equations (\ref{Eqd1})-(\ref{Eqd3}) for
various metrics have been constructed by many authors (see e.g.
\cite{CHM,Cai22,Clem,DF,DF2,SDR,SDR2}). Here we would like to
obtain the topological black hole solutions of the field equations
(\ref{Eq1})-(\ref{Eq3}) in BDM theory, by applying the conformal
transformations (\ref{conform2}) on the corresponding solutions in
the dilaton gravity theory. The $(n+1)$-dimensional topological
black hole solution of the field equations
(\ref{Eqd1})-(\ref{Eqd3}) has been obtained by one of us in
\cite{shey2} for two Liouville-type dilaton potentials
\begin{equation}\label{v2}
\bar{V}(\bar{\Phi}) = 2\Lambda_{0} e^{2\zeta_{0}\bar{\Phi}} + 2
\Lambda e^{2\zeta \bar{\Phi}},
\end{equation}
where $\Lambda_{0}$,  $\Lambda$, $ \zeta_{0}$ and $ \zeta$ are
constants. In \cite{shey2} the spacetime metric was written in the
form
\begin{equation}
d\bar{s}^2=-f(r)dt^2+\frac{dr^2}{f(r)}+r^2{R^2(r)}h_{ij}dx^{i}dx^j,
\label{met1}
\end{equation}
where $f(r)$ and $R(r)$ are functions of $r$ which should be
determined, and $h_{ij}$ is a function of coordinate $x_i$ which
spanned an $(n-1)-$dimensional hypersurface with constant scalar
curvature $(n-1)(n-2)k$. Here $k$ is a constant which
characterized the hypersurface. Without loss of generality, one
can take $k=0,1,-1$, such that the black hole horizon or
cosmological horizon in (\ref{met1}) can be a zero (flat),
positive (elliptic) or negative (hyperbolic) constant curvature
hypersurface. The Maxwell equations can be integrated immediately
to give
\begin{equation}
\bar{F}_{tr}=\frac{qe^{\frac{4\alpha\bar{\Phi}}{n-1}}}{(rR)^{n-1}},
\label{13}
\end{equation}
where $q$, an integration constant, is related to the electric
charge of black hole. Defining the electric charge via $ Q =
\frac{1}{4\pi} \int \exp\left[{-4\alpha\bar{\Phi}/(n-1)}\right]
\text{ }^{*} \bar{F} d{\Omega}, $ we get
\begin{equation}
{Q}=\frac{q\Omega _{n-1}}{4\pi},  \label{Charge}
\end{equation}
where $\Omega_{n-1}$ represents the volume of constant curvature
hypersurface described by $h_{ij}dx^idx^j$. Notice that $Q$ is
invariant under the conformal transformation (\ref{conform2}).
Using the ansatz
\begin{equation}
R(r)=e^{\frac{2\alpha\bar{\Phi}}{n-1}}, \label{ansa}
\end{equation}
one can show that the system of equations
(\ref{Eqd1})-(\ref{Eqd2}) have solutions of the form \cite{shey2}
\begin{eqnarray}
&&f(r)=-\frac{k(n-2)(\alpha^2+1)^2b^{-2\gamma}r^{2\gamma}}{(\alpha^2-1)(\alpha^2+n-2)}-
\frac{m}{r^{(n-1)(1-\gamma)-1}}+
\frac{2q^{2}(\alpha^2+1)^{2}b^{-2(n-2)\gamma}}{(n-1)(\alpha^2+n-2)}r^{2(n-2)(\gamma-1)} \nonumber \\
&+&
\frac{2\Lambda(\alpha^2+1)^{2}b^{2\gamma}}{(n-1)(\alpha^2-n)}r^{2(1-\gamma)},
\label{f1}\\
&& R(r)=\left(\frac{b}{r}\right)^{\gamma},\label{R1}\\
&&
\bar{\Phi}=\frac{(n-1)\alpha}{2(1+\alpha^2)}\ln\left(\frac{b}{r}\right),
\label{phibar1}
\end{eqnarray}
where $b$ is an arbitrary constant and $\gamma =\alpha
^{2}/(\alpha ^{2}+1)$. In the above expression, $m$ appears as an
integration constant and is related to the mass of the black hole.
In order to fully satisfy the system of equations, we must have
\cite{shey2}
\begin{equation}\label{lam}
\zeta_{0} =\frac{2}{\alpha(n-1)},   \hspace{.8cm}
\zeta=\frac{2\alpha}{n-1}, \hspace{.8cm}    \Lambda_{0} =
\frac{k(n-1)(n-2)\alpha^2 }{2b^2(\alpha^2-1)}.
\end{equation}
Notice that here  $\Lambda$ is a free parameter which plays the
role of the cosmological constant. Using the conformal
transformation (\ref{conform2}), the $(n+1)$-dimensional
topological black hole solutions of BDM theory can be obtained as
\begin{equation}
ds^2=-U(r)dt^2+\frac{dr^2}{V(r)}+r^2{H^2(r)}h_{ij}dx^{i}dx^j,
\label{met2}
\end{equation}
where $U(r)$, $V(r)$, $H(r)$ and $\Phi(r)$ are
\begin{eqnarray}
&&U(r)=-\frac{k(n-2)(\alpha^2+1)^2b^{-2\gamma(\frac{n-1}{n-3})}r^{2\gamma(\frac{n-1}{n-3})}}
{(\alpha^2-1)(\alpha^2+n-2)}-
\frac{mb^{(\frac{-4\gamma}{n-3})}}{r^{n-2}}r^{\gamma\left(n-1+\frac{4}{n-3}\right)}\nonumber \\
&&+\frac{2q^{2}(\alpha^2+1)^{2}b^{-2\gamma\left(n-2+\frac{2}{n-3}\right)}}{(n-1)\left(\alpha^2+n-2\right)r^{2[(n-2)(1-\gamma)-
\frac{2\gamma}{n-3}]}}+
\frac{2\Lambda(\alpha^2+1)^{2}b^{2\gamma(\frac{n-5}{n-3})}}{(n-1)(\alpha^2-n)}r^{2(1-\frac{\gamma(n-5)}{n-3})},
 \label{U1}\\
&&V(r)=-\frac{k(n-2)(\alpha^2+1)^2b^{-2\gamma(\frac{n-5}{n-3})}r^{2\gamma(\frac{n-5}{n-3})}}
{(\alpha^2-1)(\alpha^2+n-2)}-\frac{mb^{(\frac{4\gamma}{n-3})}}{r^{n-2}}r^{\gamma(n-1-\frac{4}{n-3})}
\nonumber \\
&&+\frac{2q^{2}(\alpha^2+1)^{2}b^{-2\gamma\left(n-2-\frac{2}{n-3}\right)}}{(n-1)(\alpha^2+n-2)r^{2[(n-2)(1-\gamma)+
\frac{2\gamma}{n-3}]}}+
\frac{2\Lambda(\alpha^2+1)^{2}b^{2\gamma(\frac{n-1}{n-3})}}{(n-1)(\alpha^2-n)}r^{2(1-\gamma(\frac{n-1}{n-3}))}
,
\label{V1}\\
&& H(r)=\left(\frac{b}{r}\right)^{\frac{(n-5)\gamma}{n-3}},
\label{H1}\\
&&\Phi(r)=\left(\frac{b}{r}\right)^{\frac{2(n-1)\gamma}{n-3}}.
\label{Phi1}
\end{eqnarray}
Applying the conformal transformation, the electromagnetic field
and the scalar potential become
\begin{eqnarray}
F_{tr}&=&\frac{qb^{(3-n)\gamma}}{r^{(n-3)(1-\gamma)+2}},
\label{Ftr2} \\
V(\Phi)&=&2\Lambda_{0}\Phi^{\frac{n(\alpha^2+1)+\alpha^2-3}{\alpha^2(n-1)}}+2\Lambda\Phi^2.
\label{V2}
\end{eqnarray}
It is worth noting that in the case $k\neq0$, these solutions are
ill-defined for the string case where $\alpha=1$ (this is
corresponding to $\omega=(n-9)/4$). It is also notable to mention
that the electric field $F_{tr}$ and the scalar field $\Phi(r)$ go
to zero as $r\rightarrow\infty$. When $\omega\rightarrow\infty$
($\alpha=0=\gamma$), these solutions reduce to
\begin{eqnarray}
U(r)=V(r)=k-\frac{m}{r^{n-2}}+\frac{2q^2}{(n-1)(n-2)r^{2(n-2)}}-\frac{2\Lambda}{n(n-1)}r^2,
\end{eqnarray}
which describes an $(n+1)$-dimensional asymptotically (anti)-de
Sitter topological black holes with a positive, zero or negative
constant curvature hypersurface (see e.g. \cite{Bril1,Cai3}). It
is easy to show that the Kretschmann scalar $R_{\mu \nu \lambda
\kappa}R^{\mu \nu \lambda \kappa}$ diverge at $r=0$ and therefore
there is an essential singularity at $r=0$. As one can see from
Eqs. (\ref{U1})-(\ref{V1}), the solutions are also ill-defined for
$\alpha=\sqrt{n}$ with $\Lambda\neq0$ (corresponding to
$\omega=-3(n+3)/4n$). The cases with $\alpha<\sqrt{n}$ and
$\alpha>\sqrt{n}$ should be considered separately. In the first
case where $ \alpha <\sqrt{n}$,  there exist a cosmological
horizon for $\Lambda >0$, while there is no cosmological horizons
if $\Lambda <0$. Indeed, in the latter case ($\alpha <\sqrt{n}$
and $\Lambda <0$) the spacetimes exhibit a variety of possible
casual structures depending on the values of the metric parameters
$\alpha $, $m$, $q$ and $k$ \cite{shey2}. Therefore, our solutions
can represent topological black hole, with inner and outer event
horizons, an extreme topological black hole, or a naked
singularity provided the parameters of the solutions are chosen
suitably. In the second case where $\alpha
>\sqrt{n}$, the spacetime has a cosmological horizon for $\Lambda <0$ despite the value of
curvature constant $k$, while for $\Lambda>0$ we have cosmological
horizon in the case $k=1$ and naked singularity for $k=0,-1$.

\section{Thermodynamics of topological BD black hole \label{IV}}
We now turn to the investigation of the thermodynamics of charged
topological BD black holes we have just found. The Hawking
temperature of the topological black hole on the outer horizon
$r_+$ can be calculated using the relation
\begin{equation}
T_{+}=\frac{\kappa}{2\pi}= \frac{U^{\text{ }^{\prime
}}(r_{+})}{4\pi\sqrt{U/V}},
\end{equation}
where $\kappa$ is the surface gravity. Then, one can easily show
that
\begin{eqnarray}\label{Tem}
T_{+}&=&-\frac{(\alpha ^2+1)}{2\pi (n-1)}\left(
\frac{k(n-2)(n-1)b^{-2\gamma}}{2(\alpha^2-1)}r_{+}^{2\gamma-1}
+\Lambda b^{2\gamma}r_{+}^{1-2\gamma}+q^{2}b^{-2(n-2)\gamma
}r_{+}^{(2n-3)(\gamma -1)-\gamma}\right)\nonumber\\
&=&-\frac{k(n-2)(\alpha ^2+1)b^{-2\gamma}}{2\pi(\alpha
^2+n-2)}r_{+}^{2\gamma-1}+\frac{(n-\alpha ^{2})m}{4\pi(\alpha
^{2}+1)}{r_{+}}^{(n-1)(\gamma -1)} \nonumber\\
&&-\frac{q^{2}(\alpha ^{2}+1)b^{-2(n-2)\gamma }}{\pi(\alpha
^{2}+n-2)}{r_{+}}^{(2n-3)(\gamma -1)-\gamma}.
\end{eqnarray}
If we compare Eq. (\ref{Tem}) with the temperature obtained in the
dilaton gravity theory \cite{shey2}, we find that the temperature
is invariant under the conformal transformation (\ref{conform2}).
This is due to the fact that the conformal parameter is regular at
the horizon. Equation (\ref{Tem}) also shows that when $k=0$, the
temperature is negative for two cases  (\emph{i}) $\alpha
>\sqrt{n}$ despite the sign of $\Lambda $, and (\emph{ii})
positive $\Lambda $ despite the value of $\alpha $. As we argued
above in these two cases we encounter cosmological horizons.
Physically it is not easy to accept the negative temperature, the
temperature on the cosmological horizon should be defined as
$T=|\kappa|/2\pi$ so that it becomes a positive since on the
cosmological horizon the surface gravity is negative.

The ADM (Arnowitt-Deser-Misner) mass $M$, entropy $S$ and electric
potential $U$ of the topological black hole can be calculated
through the use of the Euclidean action method \cite{CaiSu}. In
this approach, first the electric potential and the temperature
are fixed on a boundary with a fixed radius $r_{+}$. The Euclidean
action has two parts; bulk and surface. The first step to make the
Euclidean action is to substitute $t$ with $i\tau$. This makes the
metric positive definite:
\begin{equation}
ds^2=U(r)d\tau^2+\frac{1}{V(r)}dr^2+r^2H^2(r)h_{ij}dx^{i}dx^{j}.
\label{Eumetr}
\end{equation}
There is a conical singularity at the horizon $r=r_{+}$ in the
Euclidean metric \cite{CaiSu}. To eliminate it, the Euclidian time
$\tau$ is made periodic with period $\beta$, where $\beta$ is the
inverse of Hawking temperature. Now we obtain the Euclidean action
of $(n+1)$-dimensional Brans-Dicke-Maxwell theory. The Euclidean
action can be calculated analytically and continuously changing of
action (\ref{act1}) to Euclidean time $\tau$, i.e.,
\begin{equation}
I_{GE}=-\frac{1}{16\pi}\int_{\mathcal{M}}
d^{n+1}x\sqrt{g}\left(\Phi {R}-\frac{\omega}{\Phi}(\nabla\Phi)^2
-V(\Phi)-F_{\mu \nu}F^{\mu
\nu}\right)-\frac{1}{8\pi}\int{d^{n}x\sqrt{h}\Phi(K-K_{0})},\label{act1e}
\end{equation}
where $K_{0}$  is the trace of the extrinsic curvature on the
metric $h$ for our metric background with $q=0$  and $m=0$, which
must be added so that it can normalize the Euclidean action to
zero in this spacetime \cite{Brown}. Using the metric
(\ref{Eumetr}), we find
\begin{eqnarray}
R&=&-g^{-1/2}(g^{1/2}U^{\prime}V/U)^{\prime}-2G^{0}_{0},\label{RE}\\
K&=&-\frac{\sqrt {V} \left[rH U^{\prime}+2(n-1)\left(
UH+rUH^{\prime}\right)\right] }{2rHU}, \label{K}
\end{eqnarray} where $G^{0}_{0}$ is the (00)
component of the Einstein tensor. Inserting $U(r)$ and $V(r)$ from
(\ref{U1}) and (\ref{V1}) with $q=0$ and $m=0$ in $K$ we obtain
the extrinsic curvature for the metric background
\begin{eqnarray}\label{K0}
 K_{0}&=&
  \left(
{\frac {b}{r}} \right) ^{{\frac {2\gamma}{n-3}}}
 \left( -{\frac {k \left( n-2 \right)  \left( {\alpha}^{2}+1 \right) ^
{2}{b}^{-2\,\gamma}{r}^{2\,\gamma}}{ \left( {\alpha}^{2}-1 \right)
 \left( {\alpha}^{2}+n-2 \right) }}+{\frac {2\Lambda\, \left( {
\alpha}^{2}+1 \right) ^{2}{b}^{2\,\gamma}{r}^{2-2\,\gamma}}{
\left( n- 1 \right)  \left( {\alpha}^{2}-n \right) }}
\right)^{1/2} \nonumber \\
 && \times \left[2{b}^{2\, \gamma} \left( \alpha^2-1
\right) \left( n\gamma-n+3-5\,\gamma
 \right)  \left( {\alpha}^{2}+n-2 \right)n
\Lambda\,{r}^{2-2\,\gamma} \right. \nonumber
\\
&& \left.+\left( n-2 \right) {r}^{2\,\gamma}{b} ^{-2\,\gamma}
\left( n-{\alpha}^{2}\right) \left( n-1 \right) ^{2}k
 \left(\gamma\,n-n-6\,\gamma+3 \right)  \right]\left( n-3 \right) ^{-1
}r^{-1}  \nonumber \\
 && \times
 \left[2{b}^
{2\,\gamma}\Lambda\, \left( \alpha^2-1 \right) \left(
{\alpha}^{2}+n-2 \right) {r}^{2-2\,\gamma}+{r}^{
2\,\gamma}{b}^{-2\,\gamma}k \left( n-1 \right)  \left( n-2 \right)
 \left(n -{\alpha}^{2} \right)  \right] ^{-1}
\end{eqnarray}
Substituting Eqs. (\ref{RE})-(\ref{K0}) in action (\ref{act1e})
and using Eqs. (\ref{U1})-(\ref{V2}), after a long calculation, we
obtain the Euclidean action as
\begin{eqnarray}
I_{GE}=\beta\frac{\Omega_{n-1}}{16\pi}\left(
\frac{b^{(n-1)\gamma}(n-1)m}{(\alpha^2+1)}\right)
-\frac{\Omega_{n-1}}{4}\left(b^{(n-1)\gamma}r_{+}^{(n-1)(1-\gamma)}\right)-\beta\frac{\Omega_{n-1}q^2}{8\pi\Upsilon
r_{+}^{\Upsilon}}, \label{IE}
\end{eqnarray}
where $\Upsilon=(n-3)(1-\gamma)+1$. According to Ref.
\cite{Brown,Brown2,Brown3,Brown4}, the thermodynamical potential
can be given by $I_{GE}$, we get
\begin{equation}
I_{GE}=\beta M-S-\beta Uq, \label{GD}
\end{equation}
where $M$ is the ADM mass, $S$ and $U$ are, respectively, the
entropy and the electric potential. Comparing Eq. (\ref{IE}) with
Eq. (\ref{GD}), we find
\begin{equation}
{M}=\frac{b^{(n-1)\gamma}(n-1) \Omega _{n-1}}{16\pi(\alpha^2+1)}m,
\label{Mass}
\end{equation}
\begin{equation}
S=\frac{b^{(n-1)\gamma}r_{+}^{(n-1)(1-\gamma)}}{4}\Omega_{n-1},\label{entropy}
\end{equation}
\begin{equation}
U=\frac{qb^{(3-n)\gamma}\Omega_{n-1}}{\Upsilon
r_{+}^{\Upsilon}}.\label{elect}
\end{equation}
Comparing the conserved and thermodynamic quantities calculated in
this section with those obtained in \cite{shey2}, we find that
they are invariant under the conformal transformation
(\ref{conform2}). It is worth emphasizing that in BD theory, where
we have the additional gravitational scalar degree of freedom, the
entropy of the black hole does not follow the area law
\cite{kang}. This is due to the fact that the black hole entropy
comes from the boundary term in the Euclidean action formalism.
Nevertheless, the entropy remains unchanged under the conformal
transformations. Finally, we consider the first law of
thermodynamics for the topological black hole. It is a matter of
calculation to show that the the conserved and thermodynamic
quantities obtained above satisfy the first law of black hole
thermodynamics
\begin{equation}
dM = TdS+Ud{Q}.
\end{equation}

\section{Summary and discussion \label{V}}
To conclude, in $(n+1)$-dimensions, when the $(n-1)$-sphere of
black hole event horizn is replaced by an $(n-1)$-dimensional
hypersurface with positive, zero or negative constant curvature,
the black hole is called as a topological black hole. The
construction and analysis of these exotic black holes in anti-de
Sitter (AdS) space is a subject of much recent interest. This
interest is motivated by the correspondence between the
gravitating fields in an AdS spacetime and conformal field theory
on the boundary of the AdS spacetime. In this paper, we further
generalized these exotic solutions by constructing a class of
$(n+1)$-dimensional $(n\geq4)$ topological black holes in BDM
theory in the presence of a potential for the scalar field. In
contrast to the topological black holes in the Einstein-Maxwell
theory, which are asymptotically AdS, the topological black holes
we found here, are neither asymptotically flat nor (A)dS. Indeed,
the scalar potential plays a crucial role in the existence of
these black holes, as the negative cosmological constant does in
the Einstein-Maxwell theory. When $k=\pm 1 $, these solutions do
not exist for the string case where $\alpha=1$ (corresponding to
$\omega=(n-9)/4$). Besides they are ill-defined for
$\alpha=\sqrt{n}$ with $\Lambda\neq0$ (corresponding to
$\omega=-3(n+3)/4n$). We obtained the conserved and thermodynamic
quantities through the use of the Euclidean action method, and
verified that the conserved and thermodynamic quantities of the
solutions satisfy the first law of black hole thermodynamics. We
found that the entropy does not satisfy the area law. We also
found that the conserved and thermodynamic quantities are
invariant under the conformal transformation.

\acknowledgments{This work has been supported financially by
Research Institute for Astronomy and Astrophysics of Maragha,
Iran.}


\end{document}